\documentclass[letter,traditabstract]{aa}

\usepackage{graphicx}
\usepackage{txfonts}
\def\macc   {$\dot{M}_{\rm acc}$}
\def\mdisk {$M_{\rm disk}$}

\def\nodata {...}

\begin{document} 

 \title{Constraining disk evolution prescriptions of planet population synthesis models with observed disk masses and accretion rates}
\titlerunning{Constraining disk evolution prescriptions of planet population synthesis models}
\authorrunning{Manara, Mordasini et al.}

    \author{C.F. Manara\fnmsep\thanks{ESO Fellow}
          \inst{1}
        \and
          C. Mordasini\inst{2}
          \and
          L. Testi\inst{1,3}
        \and
          J.P. Williams\inst{4}
        \and 
         A. Miotello\inst{1}
        \and 
         G. Lodato\inst{5}
        \and
        A. Emsenhuber\inst{6,2}
          }

   \institute{European Southern Observatory, Karl-Schwarzschild-Strasse 2, 85748 Garching bei M\"unchen, Germany\\
              \email{cmanara@eso.org}
         \and
            Physikalisches Institut, University of Bern, Gesellschaftsstrasse 6 CH 3012 Bern, Switzerland
        \and
        INAF - Osservatorio Astrofisico di Arcetri, L.go E. Fermi 5, 50125 Firenze, Italy
        \and
           Institute for Astronomy, University of Hawai'i at M\"anoa, Honolulu, HI, USA
        \and
        Dipartimento di Fisica, Universit{\`a} Degli Studi di Milano, Via Celoria, 16, 20133 Milano, Italy
        \and
        Lunar and Planetary Laboratory, University of Arizona, 1629 E. University Blvd., Tucson, AZ 85721, USA
             }

   \date{Received August 9, 2019; accepted September 18, 2019}

 
  \abstract{While planets are commonly discovered around main-sequence stars, the processes leading to their formation are still far from being understood. Current planet population synthesis models, which aim to describe the planet formation process from the protoplanetary disk phase to the time exoplanets are observed, rely on prescriptions for the underlying properties of protoplanetary disks where planets form and evolve. The recent development in measuring disk masses and disk-star interaction properties, i.e., mass accretion rates, in large samples of young stellar objects demand a more careful comparison between the models and the data. We performed an initial critical assessment of the assumptions made by planet synthesis population models by looking at the relation between mass accretion rates and disk masses in the models and in the currently available data. We find that the currently used disk models predict mass accretion rate in line with what is measured, but with a much lower spread of values than observed. This difference is mainly because the models have a smaller spread of viscous timescales than what is needed to reproduce the observations. We also find an overabundance of weakly accreting disks in the models where giant planets have formed with respect to observations of typical disks. We suggest that either fewer giant planets have formed in reality or that the prescription for planet accretion predicts accretion on the planets that is too high. Finally, the comparison of the properties of transition disks with large cavities confirms that in many of these objects the observed accretion rates are higher than those predicted by the models. On the other hand, PDS70, a transition disk with two detected giant planets in the cavity, shows mass accretion rates well in line with model predictions.
}


   \maketitle
%
\section{Introduction}

While it is now well accepted that exoplanetary systems are ubiquitous, we are still debating how to explain their formation and their diversity. In particular, one of the major shortcomings in this quest is to describe correctly the properties of protoplanetary disks, the site where planets form, in the current models of planet formation \citep[e.g.,][for a review]{MR16}.

In the last decade, a big effort has been put into population synthesis models to describe what kind of exoplanetary systems are produced given some assumptions on the disk morphology and evolution, on the formation of planets, and on the accretion of material on planetesimals \citep[e.g.,][for a review]{benz14}. Both the properties of disks at the time of the formation of planets and the exact process governing the growth of dust from small grains to pebbles, and from planetesimals to planetary cores, are still flawed by several unknowns. 

In this work we attempt an initial comparison between the assumed disk structure in current planet population synthesis models \citep{mordasini12} with available observations of some of the key disk parameters, in particular the disk mass (\mdisk) and mass accretion rate onto the star (\macc). Such a comparison is the first step to validate the assumptions made by the models, on the one hand, and to identify where current models must be revised, on the other hand.

\section{Observational data}\label{sect::sample}

To date, the complete disk-bearing population of young stars in two star-forming regions with age $\sim$1-3 Myr, Lupus and Chamaeleon~I, have been observed both with optical spectroscopy with the Very Large Telecope (VLT) X-Shooter instrument and with the Atacama Large Millimeter and submillimeter Array (ALMA). These instruments currently represent the best means to measure \macc \ and \mdisk. Indeed, combining the X-Shooter data analyzed by \citet{alcala14,alcala17} and the ALMA data by \citet{ansdell16} for the targets in the Lupus complex, \citet{manara16b} showed that there is a correlation between \macc\ and \mdisk, when the latter is obtained by converting the continuum flux into dust mass. Similarly, \citet{mulders17} confirmed the \macc-\mdisk\ correlation by combining the X-Shooter data analyzed by \citet{manara16a,manara17a} with the ALMA data analyzed by \citet{pascucci16}. 

In the following, these two data sets are used as a prime comparison set for the planet population synthesis model disk parameters. We assume that the total disk mass, \mdisk, is equal to 100 times the disk dust mass, which is measured by converting the observed continuum flux assuming the opacity $\kappa_\nu = 2.3 ~ {\rm cm}^2 ~ {\rm g}^{-1} \cdot (\nu/230~ {\rm GHz})^{0.4}$ \citep{andrews13}\footnote{Only the values of \mdisk \ from \citet{ansdell16} have to be rescaled to this different opacity.} and a disk temperature of 20 K \citep{ansdell16,pascucci16,pinilla18_tds}. As in \citet{manara18b}, \mdisk \ is rescaled to the Gaia data release 2 \citep[DR2;][]{gaia,gaiadr2} distances for the individual stars, and \macc \ is also recalculated after rescaling the accretion luminosity and stellar luminosity to the Gaia DR2 distances (see Table~\ref{tab::lupus}-~\ref{tab::chaI}). 

We performed a fit of the combined sample of data in Chamaeleon~I and Lupus using these revised values. Following \citet{manara16b} and \citet{mulders17}, the fit is performed using \textsc{linmix}\footnote{https://github.com/jmeyers314/linmix} \citep{kelly07} on the objects with detected disks and measured \macc. The best fit relation is $\log\dot{M}_{\rm acc} = (0.9\pm0.1)\cdot \log M_{\rm disk} - (6.5\pm0.4)$ with a correlation coefficient $r=0.6$ and a dispersion of 0.9 dex that is slightly more dispersed and steeper than previously reported, but still within uncertainties.

To increase the sample of disks with dust cavities resolved by ALMA, the so-called transition disks, we used the compilation by \citet{pinilla18_tds} and included PDS70 \citep{keppler19,haffert19}. The list of objects considered in this work are reported in Table~\ref{tab::tds}, where \macc \ and \mdisk \ are also rescaled to the Gaia DR2 distances.

\begin{figure}[]
\centering
\includegraphics[width=0.5\textwidth]{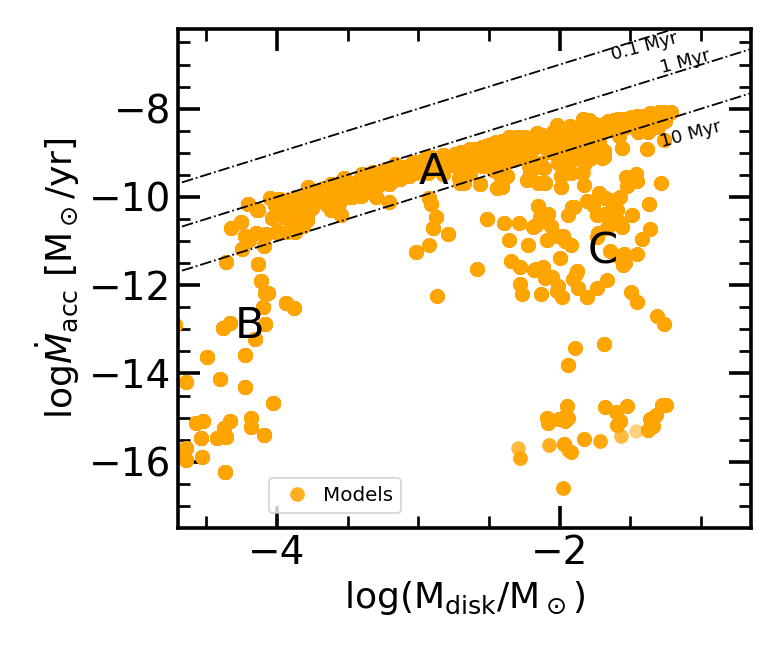}
\caption{Mass accretion rate vs. disk mass predicted by 2 Myr old disk models used in planetary synthesis population models (orange filled circles). A, B, and C show the three main loci described in the text.
     \label{fig::mdisk_macc_mod}}
\end{figure}

\section{Comparison with models for planet formation synthesis}\label{sect::comparison}
As described by \citet{mordasini09,mordasini12}, the Bern planet population synthesis models are based on the core accretion paradigm for planet formation, coupled to a model of disk evolution and tidal migration of the planets \citep{alibert05}. In particular, the disk evolution model \citep[described in][]{benz14} relies on solving the viscous evolution equation \citep{LBP74}, parameterized by an $\alpha$-parameter of 2x$10^{-3}$, coupled with a prescription for external far-ultraviolet photo-evaporation \citep{matsuyama03} with a mass loss rate randomly sampled to disperse the disk according to a typical disk lifetime \citep{haisch01}, and internal extreme-ultraviolet radiation \citep{clarke01}, which  is responsible for opening a gap in the disk when \macc$\lesssim 10^{-11}M_\odot$/yr, plus mass removal because of accretion by growing  planets. In the models the initial disk mass distribution is taken from \citet{tychoniec18} and has a mean value 30 $M_{\rm Jup}$ and a dispersion of $\sim$0.2 dex. The initial radii distribution is set using the relation between disk mass and disk characteristic radius ($R_{\rm c}$) described by \citet{andrews10}, assuming this is valid for the initial disk masses. We note, however, that this relation is based on many disks showing substructures, which are known to be the largest \citep{long19}, and on evolved disks, whose sizes could not reflect the initial size distribution, but could be the effect of radial drift in the disk \citep[e.g.,][]{rosotti19}.
The models discussed in this work assume central stars with a mass of 1 $M_\odot$. This assumption is only relevant for the following discussion as a second order correction. Indeed, it is known that the disk mass depends on the stellar mass \citep[e.g.,][]{ansdell16,pascucci16}, but the disk masses covered by models reproduce the full range of observed disk masses for disks around a large range of stellar mass. 
The information on the mass of the central star enters only indirectly in the values of the viscosity ($\nu$) used in the models. Indeed, this parameter is expressed as $\nu = \alpha c_s H$, where $c_s$ is the sound speed and $H$ the scale height of the disk; the latter two parameters are obtained by solving for the vertical structure equilibrium due to viscous heating and stellar irradiation as described by \citet{CG97}. Similarly, a dependence of the mass loss rate from photo-evaporation with stellar mass is expected \citep[e.g.,][]{owen11}, but as a second order effect.

In the following, we use the snapshot of the models at $t$ = 2 Myr for a comparison with the data. The models we considered start with 100 seeds of planetary systems, which is the most consistent value with planet detections with \textit{Kepler} and HARPS \citep{mulders19}. 
The age at which the models are evaluated is chosen to be in line with typical estimates for the ages of the Chamaeleon~I and Lupus regions, which are considered to be $\sim$1-3 Myr old. At this age, $\lesssim$10\% of the modeled disks have masses below the numerical minimum density, i.e., they have dissipated. 

\begin{figure}[]
\centering
\includegraphics[width=0.5\textwidth]{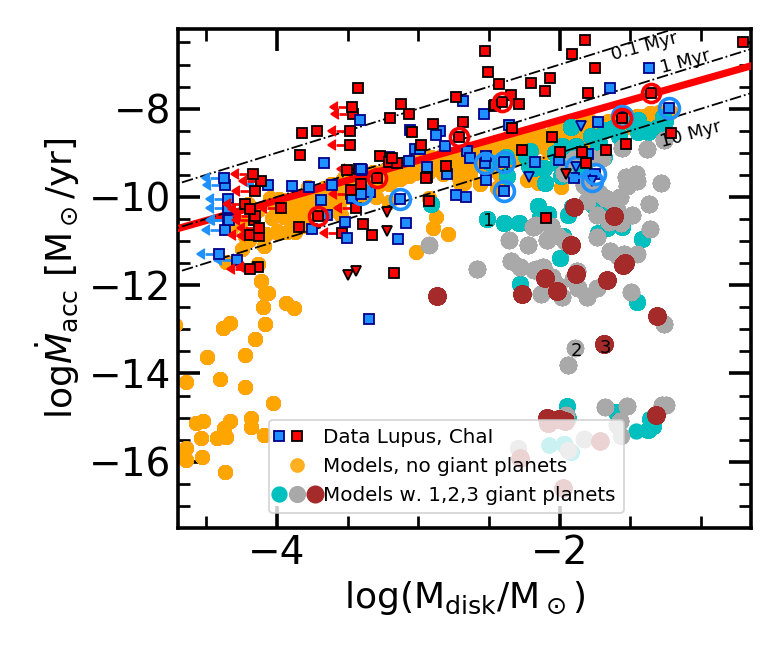}
\caption{Mass accretion rate vs. disk mass for the models (filled circles) and for the observed population of disks in the Lupus (blue symbols) and Chamaeleon~I (red symbols) regions. Squares are used for disks with measured disk mass and mass accretion rates; squares with arrow report upper limits on the disk mass; downward facing triangles report dubious accretors, i.e., objects with mass accretion rates compatible with chromospheric emission; and the transition disks are highlighted with a circle around the symbol. The red line indicates the best fit. The models are colored for the number of giant planets ($M_{\rm planet} > 300 M_\oplus$) in the system: orange for no giant planets, cyan for one giant
planet, gray for two giant
planets, and brown for three giant planets. 
     \label{fig::mdisk_macc_mod_planets}}
\end{figure}

The distribution of \macc \ and \mdisk \ for the models (Fig.~\ref{fig::mdisk_macc_mod}) presents three main loci. First (A), $\sim$65\% of the model points are located along a major \macc-\mdisk \ sequence, almost parallel to lines of constant \mdisk/\macc, and between the lines of \macc-\mdisk=1 Myr and 10 Myr. Second (B), $\sim$5\% of the points are found at \mdisk $\lesssim10^{-4} M_\odot$ and \macc $\lesssim 10^{-11} M_\odot$/yr, to the bottom left of the plot. Third (C), $\sim$20\% of the points are located at $10^{-3} M_\odot \lesssim$ \mdisk $\lesssim10^{-1} M_\odot$ and at \macc \ lower than the typical values found in models in the same \mdisk \ range. 
These three main loci are easily understood as: (A) the main location where disks in the models spend their lifetime; (B) the locus of the disks in which internal photo-evaporation has overcome the effect of viscous evolution, has stopped accretion, and is rapidly dissipating the disks; and (C) the locus where giant planet formation has taken place, respectively. 
We note that the photo-evaporation prescription used in the models is directly responsible for the number of objects present in locus (B) and for the values of \mdisk \ and \macc \ , where the separation between loci (A) and (B) is visible. A more vigorous mass loss rate, such as that produced by X-ray photo-evaporation \citep[e.g.,][]{picogna19}, would change the shape of the locus (B) and would accelerate the disk dispersal. Thus, a more vigorous mass loss rate would increase the fraction of models in locus (B).
In the disk models in locus (C), a large percentage of the disk accretion flow ends up on the accreting planets instead of the star. 
Indeed, the models in which at least one giant planet -- defined as having $M_{\rm planet} > 1\,M_{\rm Jup}$ -- are highlighted in Fig.~\ref{fig::mdisk_macc_mod_planets} and are located in locus (C), represent $\sim$75\% of the models in this locus. As expected, we find the accreting giant planets in massive disk with low or no accretion \citep[e.g.,][]{WC11,rosotti17}.

The comparison between the data and the models of Fig.~\ref{fig::mdisk_macc_mod_planets} shows that the model parameters have some similarities to the observations. In particular, the main locus (A) of the \macc-\mdisk \ values of the models is in line with the data. Most notably, in the disk mass range from $10^{-4} - 10^{-2} M_\sun$ the upper bound of the locus of the models follows very well the best fit of the observations, whereas at \mdisk $\gtrsim10^{-2} M_\sun$ the models in the main group (A) tend to bend to slightly lower \macc. Moreover, the typical values of \macc \ are within the observed values for the Chamaeleon~I and Lupus regions. 

However, a number of significant differences are present. 
The dispersion of \macc \ at any \mdisk \ of the main locus of the models ($\sim$0.3 dex) is much smaller than that of the data ($\sim$0.9 dex). This is partially because the models shown in this work are not convolved with the typical observational uncertainties ($\lesssim$0.4 dex). However, the discrepancy is larger than this effect. 
This discrepancy is in line with the results of \citet{lodato17} and \citet{mulders17}, who postulated a large spread of model parameters, in particular of the viscous timescale and/or $\alpha$, and a long viscous timescale on the order of $\sim$1 Myr to reproduce the observed spread. More specifically, \citet{mulders17} require that the values of $\alpha$ are distributed around a typical value of $10^{-3}$ with a dispersion of 2 dex to reproduce the observations in contrast to the single value assumed in the models. The single value of $\alpha$ however still connects to a dispersion in viscous timescales ($t_\nu$) since the initial disk radii present a distribution of values. Indeed, $t_\nu \propto R_0^{3/2}(2-\gamma)^{-2} \alpha^{-1} (H/R)_0^{-2}$, thus we can assume a value of $\gamma=1.5$, $H/R=0.1$, and $\alpha=2\cdot 10^{-3}$ to derive the viscous timescales of the models. This distribution has a spread of $\lesssim0.5$ dex, smaller than the spread of $\sim1$ dex needed by \citet{lodato17} to reproduce the observed \mdisk-\macc. 
The values of \macc \ for the main locus of the models are then within the typical observed values, but are systematically below the median of the distribution, i.e., the best fit. This fact is related to the assumed value of $\alpha$ and to the other disk initial parameters. A higher value of $\alpha$ increases the predicted \macc, but implies a shorter timescale of the disks. 

Finally, almost no overlap is present between the observed data and the model points in the (B) photo-evaporative and (C) giant planet forming disk regions. On the one hand, the fact that we do not observe the photo-evaporative disks is easily explainable. These disks are predicted to have low mass, at \mdisk \ values where ALMA surveys are incomplete and dominated by upper limits \citep[e.g.,][]{ansdell16,pascucci16}, and the lifetime of these disks is expected to be very short, i.e., $\sim 10^5$ yr \citep[e.g.,][]{EP17}. Furthermore, the values of \macc \ reported by the models for these disks are well below the lowest values detectable from spectra of accreting young stars \citep[e.g.,][]{manara13a}. These predicted disks could be Class~III, i.e., diskless, young stellar objects.

On the other hand, it is worth asking whether the number of models in the giant planet forming disk locus is in line with observations. At the values of \mdisk \ corresponding to these models the ALMA surveys are complete, since any disk that shows an infrared excess with Spitzer has been targeted and the sensitivity of the surveys is always such that these massive disks are readily detected. Even in the case in which the disk surveys were not complete, there should be no bias against massive disks with already formed giant planets. Similarly, the spectroscopic surveys connected to the ALMA surveys are $\sim$ 95\% complete, and they are usually slightly incomplete in the lower stellar mass end of the distribution of targets, which corresponds to the lower disk masses. Therefore, it is safe to assume that both the massive disks and the corresponding stars have been observed in the ALMA and X-Shooter surveys. It is in any case possible that the observed targets have values of \macc \ lower than what is detectable with X-Shooter spectra. As discussed by \citet{manara13a} and \citet{ingleby11}, among others, \macc compatible with or lower than the typical chromospheric noise of young stellar objects are not measurable from near-ultraviolet and optical spectra. This limit depends on the stellar mass and is typically $\sim 10^{-11}-10^{-10} M_\odot$/yr, exactly in the region where the giant planet forming disks with higher \macc \ are located. Both \citet{alcala14,alcala17} and \citet{manara16a,manara17a} have reported a number of objects present in the surveys of Lupus and Chamaeleon~I for which the excess emission in the spectra has a strength that is compatible with being chromospheric. These objects, highlighted with downward triangles in the plots and referred to as ``weak-accretors'', are the only candidates to have real \macc \ lower than this chromospheric noise. However, these weak-accretors account for only $\sim 6-12$ \% of the observed population of objects with a disk in these two regions, and have in some cases \mdisk \ lower than the lowest masses of giant planet forming disks in the models. Even in the case that these are all objects whose \macc \ is in line with that reported for giant planet forming disks, the fraction of disks with these low \macc \ is smaller by at least a factor of two than predicted by current planet synthesis population models. This is well in line with the overabundance of planets formed in models in comparison with current detections of planets with Kepler and HARPS \citep{mulders19}.

\section{Transition disks and disks with planets}\label{sect::tds}

\begin{figure}[]
\centering
\includegraphics[width=0.5\textwidth]{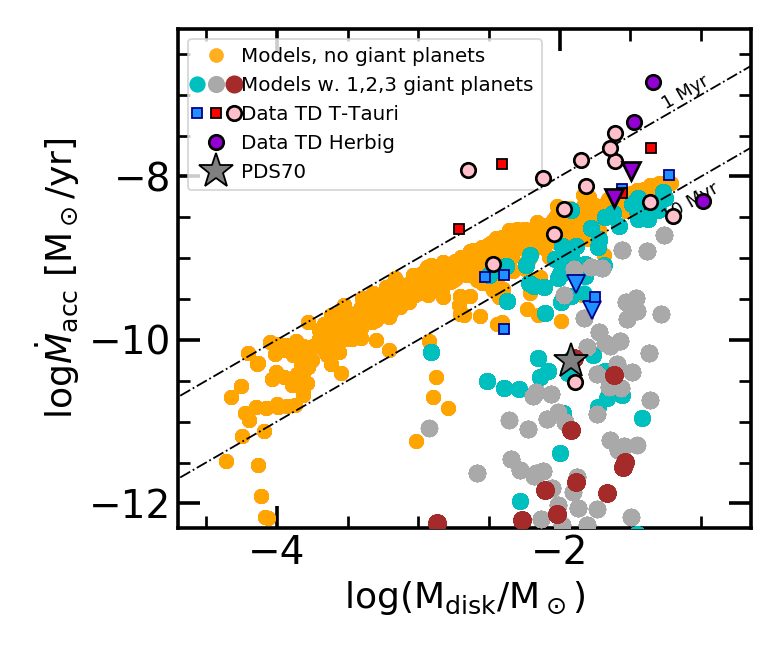}
\caption{Mass accretion rate vs. disk mass for the models (filled circles, colors as in Fig.~\ref{fig::mdisk_macc_mod_planets}) and for transition disks with resolved cavities. The pink symbols show transition disks around T Tauri stars and the violet symbols the transition disks around Herbig stars from \citet{pinilla18_tds}. The gray star symbol refers to PDS~70. Other symbols as in Fig.~\ref{fig::mdisk_macc_mod_planets}.
     \label{fig::mdisk_macc_mod_tds}}
\end{figure}

To better understand the population of giant planets forming disks predicted by the models we compiled a list of known transition disks with dust cavities resolved by ALMA observations, i.e., larger than $\sim$20 -- 30 au \citep[][see Table~\ref{tab::tds}]{pinilla18_tds}. One of the favored explanations for the cavities observed with ALMA is indeed the presence of one, or more, giant planets in the inner regions of these disks. In one case, PDS~70, two accreting giant planets have been detected in the dust depleted cavity \citep{keppler18,haffert19}. The disk masses and \macc \ for these targets are shown in Fig.~\ref{fig::mdisk_macc_mod_tds} together with the transition disks located in the Lupus and Chamaeleon~I regions and the models. All the transition disks with \mdisk$>10^{-3} M_\odot$ have been resolved with ALMA. When the central star is not a Herbig star, their \macc \ are within the range of the models given their \mdisk \ in $\sim$80\% of the cases and are always compatible within the uncertainties. However, $\sim$10--20\% of the targets are only compatible with models of disks with no yet formed giant planets. In a way, this is similar to what was discussed by \citet[][and references therein]{EP17}, meaning that there are too many transition disks with high accretion rates than predicted by photo-evaporation models. Allowing a more rapid dispersal of the disk due to planet-induced photo-evaporation \citep{rosotti13} might mitigate the discrepancy here by allowing a faster disk dispersal in the low-\macc \ and high-\mdisk \ objects that formed a planet predicted by the models discussed here. A similar effect would also be obtained including stronger photo-evaporative winds, as in the case of X-ray photo-evaporation, although even these models still are unable to reproduce the observed accreting transition disks with large cavities \citep{picogna19}.

It is worth noting that five transition disks around T Tauri stars ($\sim$25\%) present \macc \ values compatible with those of models in which at least two giant planets have formed, either measured \macc \ values or because they are weak-accretors. In particular, PDS~70 falls well within the region where models predict two giant planets to have formed, and J1604--2165, another well-studied transition disk, is also in the same region of the parameter space. Furthermore, two of the five Herbig stars with transition disks reported in this work only have an upper limit on the value of \macc, and they can potentially be compatible with having \macc \ in line with disks with giant planets.

\section{Conclusions}

We performed the first comparison between observed properties of disks, namely their mass and mass accretion rate on the central star, with disk properties predicted by models adopted for planetary synthesis population studies for $\sim$1-3 Myr old protoplanetary disks. 

We showed that the planetary synthesis population models typically predict disks with lower \macc \ than the median measured values, but still within the observed spread. However, the spread of \macc \ predicted by these disk models is too small to match the observed spread, since the spread in the viscous timescale is too small. This is in line with what was suggested by \citet{lodato17} and \citet{mulders17}. Therefore, planetary synthesis population models must use a larger dispersion of viscous timescales to match the observations.

The planetary synthesis models discussed in this work predict a larger percentage, of $\sim$20\%, of disks with very low \macc$\lesssim 10^{-10} M_\odot$/yr and high \mdisk$\gtrsim 3\cdot 10^{-3} M_\odot$ than what is observed in $\sim$1-3 Myr old disk populations, i.e., $\sim$6--12\%. This discrepancy points to either the fact that fewer giant planets are forming in disks than what is predicted by models, as pointed out also by \citet{mulders19}, or to the fact that the current prescription of gas accretion onto planets overpredicts the real accretion rate onto planets. The latter would make the accretion rate onto the star lower than observed. This might be related to the models that underpredict the number of intermediate mass planets when compared to the planetary mass function deduced from microlensing surveys \citep{suzuki18}.

The comparison between the models and the measured values of \macc \ and \mdisk \ for transition disks with large cavities, which are possibly explained by the presence of giant planets, shows some agreement with this hypothesis that the cavities are carved by giant planets. Indeed, the majority ($\sim$80\%) of transition disks have values of \macc \ and \mdisk that are\ compatible with what is expected for disks with at least one giant planets formed. Most notably, the system PDS~70 has measured \mdisk \ and \macc \ well in line with predictions for systems with two giant planets, which have been observed in this system. However, there is a small percentage, $\sim$20\%, of transition disks with \macc \ that is higher than the highest \macc \ predicted by the models. Different initial conditions for viscously evolving disks are needed, or different models of disk evolution should be explored, such as magnetic disk wind driven evolution, to explain these objects.

Future work should focus on detailed comparisons between the models and the current and future observations. In particular, it is important to test whether the disks with \mdisk$\lesssim 10^{-4} M_\odot$ and \macc$\lesssim 10^{-12} M_\odot$/yr, predicted by the currently adopted prescriptions for photo-evaporation and by planet synthesis population models, can be observed. To this aim, higher sensitivity and resolution ALMA surveys are needed. Related to this point, a more detailed description of the effect of internal photo-evaporation, for example including X-ray photo-evaporation \citep[e.g.,][]{picogna19}, should be explored in the planet synthesis population models to understand how the picture of disk properties and planet formation would be affected. Finally, this work did not discuss how these properties vary with the assumed stellar masses and at later times. This must be the subject of future studies.

\begin{acknowledgements}
We are grateful to the referee for a constructive report that helped us to improve the manuscript. 
We thank Paola Pinilla for sharing information on the transition disk masses and for insightful comments.
CFM acknowledges support through the ESO fellowship. C.M. acknowledges the support from the Swiss National Science Foundation under grant BSSGI0$\_$155816 ``PlanetsInTime''. 
This project has received funding from the European Union’s Horizon 2020 research and innovation programme under the Marie Sklodowska-Curie grant agreement No 823823 (DUSTBUSTERS).
This work was partly supported by the Deutsche Forschungs-Gemeinschaft (DFG, German Research Foundation) - Ref no. FOR 2634/1 TE 1024/1-1. 
This work made use of the Python packages Numpy and matplotlib. 
This work has made use of data from the European Space Agency (ESA) mission {\it Gaia} (\url{https://www.cosmos.esa.int/gaia}), processed by the {\it Gaia}
Data Processing and Analysis Consortium (DPAC, \url{https://www.cosmos.esa.int/web/gaia/dpac/consortium}). Funding for the DPAC has been provided by national institutions, in particular the institutions
participating in the {\it Gaia} Multilateral Agreement.
\end{acknowledgements}

%

\begin{thebibliography}{}
\bibitem[Alcal{\'a} et al.(2014)]{alcala14} Alcal{\'a}, J.~M., Natta, A., Manara, C.~F. et al.\ 2014, \aap, 561, A2.

\bibitem[Alcal{\'a} et al.(2017)]{alcala17} Alcal{\'a}, J.~M., Manara, C.~F., Natta, A. et al.\ 2017, \aap, 600, A20.

\bibitem[Alibert et al.(2005)]{alibert05} Alibert, Y., Mordasini, C., Benz, W., et al.\ 2005, \aap, 434, 343

\bibitem[Andrews et al.(2010)]{andrews10} Andrews, S.~M., Wilner, D.~J., Hughes, A.~M., et al.\ 2010, \apj, 723, 1241

\bibitem[Andrews et al.(2013)]{andrews13} Andrews, S.~M., Rosenfeld, K.~A., Kraus, A.~L. et al.\ 2013, \apj, 771, 129.

\bibitem[Ansdell et al.(2016)]{ansdell16} Ansdell, M., Williams, J.~P., van der Marel, N. et al.\ 2016, \apj, 828, 46 


\bibitem[Benz et al.(2014)]{benz14} Benz, W., Ida, S., Alibert, Y., et al.\ 2014, Protostars and Planets VI, 691

\bibitem[Chiang \& Goldreich(1997)]{CG97} Chiang, E.~I., \& Goldreich, P.\ 1997, \apj, 490, 368


\bibitem[Clarke et al.(2001)]{clarke01} Clarke, C.~J., Gendrin, A., \& Sotomayor, M.\ 2001, \mnras, 328, 485

\bibitem[Ercolano \& Pascucci(2017)]{EP17} Ercolano, B. \& Pascucci, I.\ 2017, Royal Society Open Science, 4, 170114

\bibitem[Fairlamb et al.(2015)]{fairlamb15} Fairlamb, J.~R., Oudmaijer, R.~D., Mendigut{\'\i}a, I., et al.\ 2015, \mnras, 453, 976


\bibitem[Gaia Collaboration et al.(2016)]{gaia} Gaia Collaboration, Prusti, T., de Bruijne, J.~H.~J. et al.\ 2016, \aap, 595, A1 

\bibitem[Gaia Collaboration et al.(2018)]{gaiadr2} Gaia Collaboration, Brown, A.~G.~A., Vallenari, A. et al.\ 2018, arXiv:1804.09365 

\bibitem[Haffert et al.(2019)]{haffert19} Haffert, S.~Y., Bohn, A.~J., de Boer, J., et al.\ 2019, Nature Astronomy, 329

\bibitem[Haisch et al.(2001)]{haisch01} Haisch, K.~E., Lada, E.~A., \& Lada, C.~J.\ 2001, \apjl, 553, L153


\bibitem[Ingleby et al.(2011)]{ingleby11} Ingleby, L., Calvet, N., Bergin, E., et al.\ 2011, \apj, 743, 105

\bibitem[Kelly(2007)]{kelly07} Kelly, B.~C.\ 2007, \apj, 665, 1489 

\bibitem[Keppler et al.(2018)]{keppler18} Keppler, M., Benisty, M., M{\"u}ller, A., et al.\ 2018, \aap, 617, A44

\bibitem[Keppler et al.(2019)]{keppler19} Keppler, M., Teague, R., Bae, J., et al.\ 2019, \aap, 625, A118

\bibitem[Kudo et al.(2018)]{kudo18} Kudo, T., Hashimoto, J., Muto, T., et al.\ 2018, \apjl, 868, L5

\bibitem[Lynden-Bell \& Pringle(1974)]{LBP74} Lynden-Bell, D., \& Pringle, J.~E.\ 1974, \mnras, 168, 603

\bibitem[Lodato et al.(2017)]{lodato17} Lodato, G., Scardoni, C.~E., Manara, C.~F. et al.\ 2017, \mnras, 472, 4700.

\bibitem[Long et al.(2019)]{long19} Long, F., Herczeg, G.~J., Harsono, D., et al.\ 2019, \apj, 882, 49

\bibitem[Manara et al.(2013)]{manara13a} Manara, C.~F., Testi, L., Rigliaco, E., et al.\ 2013, \aap, 551, A107

\bibitem[Manara et al.(2014)]{manara14} Manara, C.~F., Testi, L., Natta, A., et al.\ 2014, \aap, 568, A18

\bibitem[Manara et al.(2016a)]{manara16a} Manara, C.~F., Fedele, D., Herczeg, G.~J. et al.\ 2016a, \aap, 585, A136.

\bibitem[Manara et al.(2016b)]{manara16b} Manara, C.~F., Rosotti, G., Testi, L. et al.\ 2016b, \aap, 591, L3 

\bibitem[Manara et al.(2017)]{manara17a} Manara, C.~F., Testi, L., Herczeg, G.~J. et al.\ 2017, \aap, 604, A127.

\bibitem[Manara et al.(2018)]{manara18b} Manara, C.~F., Morbidelli, A., \& Guillot, T.\ 2018, \aap, 618, L3

\bibitem[Matsuyama et al.(2003)]{matsuyama03} Matsuyama, I., Johnstone, D., \& Hartmann, L.\ 2003, \apj, 582, 893

\bibitem[Mendigut{\'\i}a et al.(2011)]{mendigutia11} Mendigut{\'\i}a, I., Calvet, N., Montesinos, B., et al.\ 2011, \aap, 535, A99

\bibitem[Morbidelli \& Raymond(2016)]{MR16} Morbidelli, A., \& Raymond, S.~N.\ 2016, Journal of Geophysical Research (Planets), 121, 1962

\bibitem[Mordasini et al.(2009)]{mordasini09} Mordasini, C., Alibert, Y., \& Benz, W.\ 2009, \aap, 501, 1139


\bibitem[Mordasini et al.(2012)]{mordasini12} Mordasini, C., Alibert, Y., Klahr, H., et al.\ 2012, \aap, 547, A111


\bibitem[Mulders et al.(2017)]{mulders17} Mulders, G.~D., Pascucci, I., Manara, C.~F., et al.\ 2017, \apj, 847, 31

\bibitem[Mulders et al.(2019)]{mulders19} Mulders, G.~D., Mordasini, C., Pascucci, I., et al.\ 2019, arXiv e-prints, arXiv:1905.08804

\bibitem[Natta et al.(2006)]{natta06} Natta, A., Testi, L., \& Randich, S.\ 2006, \aap, 452, 245

\bibitem[Owen et al.(2011)]{owen11} Owen, J.~E., Ercolano, B., \& Clarke, C.~J.\ 2011, \mnras, 412, 13

\bibitem[Pascucci et al.(2016)]{pascucci16} Pascucci, I., Testi, L., Herczeg, G.~J. et al.\ 2016, \apj, 831, 125 

\bibitem[Picogna et al.(2019)]{picogna19} Picogna, G., Ercolano, B., Owen, J.~E., et al.\ 2019, \mnras, 487, 691

\bibitem[Pinilla et al.(2018)]{pinilla18_tds} Pinilla, P., Tazzari, M., Pascucci, I., et al.\ 2018, \apj, 859, 32

\bibitem[Rigliaco et al.(2015)]{rigliaco15} Rigliaco, E., Pascucci, I., Duchene, G., et al.\ 2015, \apj, 801, 31

\bibitem[Rosotti et al.(2013)]{rosotti13} Rosotti, G.~P., Ercolano, B., Owen, J.~E., et al.\ 2013, \mnras, 430, 1392

\bibitem[Rosotti et al.(2017)]{rosotti17} Rosotti, G.~P., Clarke, C.~J., Manara, C.~F., \& Facchini, S.\ 2017, \mnras, 468, 1631 

\bibitem[Rosotti et al.(2019)]{rosotti19} Rosotti, G.~P., Tazzari, M., Booth, R.~A., et al.\ 2019, \mnras, 486, 4829

\bibitem[Schisano et al.(2009)]{schisano09} Schisano, E., Covino, E., Alcal{\'a}, J.~M., et al.\ 2009, \aap, 501, 1013

\bibitem[Suzuki et al.(2018)]{suzuki18} Suzuki, D., Bennett, D.~P., Ida, S., et al.\ 2018, \apjl, 869, L34

\bibitem[Tychoniec et al.(2018)]{tychoniec18} Tychoniec, {\L}., Tobin, J.~J., Karska, A., et al.\ 2018, \apjs, 238, 19


\bibitem[Williams \& Cieza(2011)]{WC11} Williams, J.~P., \& Cieza, L.~A.\ 2011, \araa, 49, 67





\end{thebibliography}
%

\appendix

\section{Observational data used in the paper}
We report the values of \macc \ and $M_{\rm dust}$ used in this work. The latter is converted to \mdisk \ assuming a gas-to-dust ratio of 100. All the values have been rescaled with respect to their original papers using the more recent Gaia DR2 distances \citep{gaiadr2}. The information for the targets in the Lupus region (Table~\ref{tab::lupus}) are taken from \citet{alcala14,alcala17} for the accretion properties, and from \citet{ansdell16} for the disk masses. As discussed in the text, the latter are rescaled to the same opacities used in other works \citep{andrews13}. The accretion parameters for the Chamaeleon~I targets are taken from \citet{manara16a,manara17a}, while the disk masses are taken from \citet{pascucci16}; these are reported in Table~\ref{tab::chaI}. 

The properties for the transition disks (Table~\ref{tab::tds}) mainly come from the compilation of \citet{pinilla18_tds}; the references for the accretion rates are reported in the table. In addition to that, accretion rates and disk masses for PDS70 are reported \citep{keppler18,haffert19}; those for HD142666 are reported as well. 

\begin{table*}  
\begin{center} 
\footnotesize 
\caption{\label{tab::lupus} Stellar and disk masses for the Lupus targets used  } 
\begin{tabular}{l| ccc|l   } 
\hline \hline 
Name   &        dist  &  $\log\dot{M}_{\rm acc}$ & $M_{\rm dust}$       &         Notes \\ 
 &                         [pc]  &   [$M_\odot$/yr]             & [$M_\oplus$] & \\
\hline
Sz65                   & 155  &  -9.54   &   20.30    &  na \\
Sz66                   & 157  &  -8.50   &   4.78    &  \nodata \\
J15450887-3417333      & 154  &  -8.36   &   14.51    &  \nodata \\
Sz68                   & 154  &  -8.40   &   46.68    &  na \\
Sz69                   & 154  &  -9.48   &   5.29    &  \nodata \\
Sz71                   & 155  &  -9.02   &   52.68    &  \nodata \\
Sz72                   & 155  &  -8.60   &   4.47    &  \nodata \\
Sz73                   & 156  &  -8.12   &   9.77    &  \nodata \\
Sz74                   & 158  &  -7.83   &   6.87    &  \nodata \\
Sz81A                  & 159  &  -8.92   &   3.16    &  \nodata \\
Sz82                   & 158  &  -7.98   &   196.68   &  td \\
Sz83                   & 159  &  -7.08   &   141.92    &  \nodata \\
Sz84                   & 152  &  -9.23   &   9.93    &  td \\
Sz129                  & 161  &  -8.30   &   61.82    &  \nodata \\
J15592523-4235066      & 147  &  -11.29   &  $<$0.05   & \nodata \\
RYLup                  & 159  &  -8.16   &   91.05    &  td \\
J16000060-4221567      & 161  &  -9.73   &   0.81    &  \nodata \\
J16000236-4222145      & 164  &  -9.56   &   42.17    &  \nodata \\
J16002612-4153553      & 164  &  -9.76   &   0.42    &  \nodata \\
Sz130                  & 160  &  -9.09   &   2.08    &  \nodata \\
MYLup                  & 156  &  -9.63   &   56.60    &  td,na \\
Sz131                  & 160  &  -9.18   &   2.88    &  \nodata \\
Sz133                  & 153  &  -99.00   &  21.13    &  sl \\
Sz88A                  & 158  &  -8.49   &   2.93    &  \nodata \\
Sz88B                  & 159  &  -10.05   &  $<$0.06   & \nodata \\
J16070384-3911113      & 158  &  -12.76   &  1.48    &  sl \\
Sz90                   & 160  &  -8.96   &   7.33    &  \nodata \\
J16073773-3921388      & 174  &  -10.40   &  0.76    &  \nodata \\
Sz95                   & 158  &  -9.40   &   1.33    &  \nodata \\
J16080017-3902595      & 159  &  -10.56   &  1.00    &  \nodata \\
Sz96                   & 156  &  -9.37   &   1.31    &  \nodata \\
J16081497-3857145      & 158  &  -10.60   &  2.73    &  \nodata \\
Sz97                   & 157  &  -9.88   &   1.51    &  \nodata \\
Sz98                   & 156  &  -7.54   &   75.61    &  \nodata \\
Sz99                   & 159  &  -9.73   &   $<$0.06   & \nodata \\
Sz100                  & 136  &  -9.87   &   13.43    &  td \\
J160828.1-391310       & 175  &  -11.42   &  $<$0.07   & na \\
Sz103                  & 159  &  -9.33   &   3.83    &  \nodata \\
J16083070-3828268      & 156  &  -9.32   &   43.06    &  td,na \\
Sz104                  & 165  &  -10.03   &  1.13    &  \nodata \\
V856Sco                & 161  &  -99.00   &  18.89    &  1 \\
Sz106                  & 161  &  -10.07   &  0.68    &  sl \\
Sz108B                 & 168  &  -9.62   &   9.98    &  \nodata \\
J16084940-3905393      & 159  &  -9.77   &   0.56    &  \nodata \\
V1192Sco               & 150  &  -99.00   &  0.27    &  sl \\
Sz110                  & 159  &  -8.84   &   5.12    &  \nodata \\
J16085324-3914401      & 167  &  -10.00   &  7.19    &  \nodata \\
J16085373-3914367      & 158  &  -10.94   &  1.04    &  \nodata \\
Sz111                  & 158  &  -9.47   &   58.71    &  td \\
J16085529-3848481      & 157  &  -10.72   &  0.59    &  \nodata \\
Sz112                  & 160  &  -9.94   &   1.30    &  td \\
Sz113                  & 163  &  -9.12   &   7.78    &  \nodata \\
J16090141-3925119      & 164  &  -9.95   &   6.17    &  \nodata \\
Sz114                  & 162  &  -9.17   &   33.14    &  \nodata \\
Sz115                  & 157  &  -9.57   &   $<$0.06   & \nodata \\
J16092697-3836269      & 159  &  -8.25   &   1.29    &  \nodata \\
Sz117                  & 158  &  -8.91   &   3.44    &  \nodata \\
Sz118                  & 163  &  -9.21   &   22.22    &  \nodata \\
J16095628-3859518      & 156  &  -10.96   &  2.39    &  \nodata \\
J16100133-3906449      & 192  &  -9.74   &   $<$0.12   & \nodata \\
J16101857-3836125      & 158  &  -10.76   &  $<$0.06   & \nodata \\
J16101984-3836065      & 158  &  -10.52   &  $<$0.06   & \nodata \\
J16102955-3922144      & 163  &  -10.05   &  2.48    &  td \\
Sz123B                 & 158  &  -10.24   &  $<$0.06   & sl \\
Sz123A                 & 158  &  -9.21   &   13.33    &  td \\
J16115979-3823383      & 164  &  -10.53   &  $<$0.06   & \nodata \\
J16124373-3815031      & 159  &  -9.07   &   9.96    &  \nodata \\
J16134410-3736462      & 160  &  -9.24   &   0.72    &  \nodata \\
\hline 
\end{tabular} 
\tablefoot{Stellar properties adapted from \citet{alcala14,alcala17} and disk masses from \citet{ansdell16} (rescaled for different opacity) using the Gaia DR2 \citep{gaiadr2} distances. Notes: na = non-accretor, sl=sub-luminous, td = transition disk. } 
\end{center} 
\end{table*}

\begin{table*}  
\begin{center} 
\footnotesize 
\caption{\label{tab::chaI} Stellar and disk masses for the Chamaeleon~I targets used } 
\begin{tabular}{l|l| ccc|l   } 
\hline \hline 
2MASS name   &   Other names    & dist    &    $\log\dot{M}_{\rm acc}$  &         $M_{\rm dust}$   &      Notes \\
 &            &             [pc]  &   [$M_\odot$/yr]            & [$M_\oplus$] & \\
\hline
2MASSJ10533978-7712338  &  2MJ10533978-7712338     &  191.81  &  -11.72     & 2.2259      &   sl \\
2MASSJ10555973-7724399  &  T3                      &  185.08  &  -8.43      & 15.3631     &   \nodata \\
2MASSJ10561638-7630530  &  ESOHalpha553            &  196.48  &  -10.78     & 1.9879      &   na \\
2MASSJ10563044-7711393  &  T4                      &  183.09  &  -9.22      & 50.9436     &   \nodata \\
2MASSJ10574219-7659356  &  T5                      &  190.00  &  -8.35      & 4.2586      &   \nodata \\
2MASSJ10580597-7711501  &  2MJ10580597-7711501     &  186.57  &  -11.68     & 1.2118      &   na \\
2MASSJ10581677-7717170  &  Sz-Cha                  &  189.84  &  -7.65      & 147.4129    &   td \\
2MASSJ10590108-7722407  &  TW-Cha                  &  185.21  &  -8.65      & 29.3149     &   \nodata \\
2MASSJ10590699-7701404  &  CR-Cha                  &  187.48  &  -8.55      & 203.0813    &   \nodata \\
2MASSJ11004022-7619280  &  T10                     &  191.54  &  -8.98      & 33.5953     &   \nodata \\
2MASSJ11022491-7733357  &  CS-Cha                  &  176.26  &  -8.20      & 92.0593     &   td \\
2MASSJ11023265-7729129  &  CHXR71                  &  195.35  &  -10.24     & $<$ 0.4043  &   na \\
2MASSJ11025504-7721508  &  T12                     &  182.24  &  -8.54      & 0.5047      &   \nodata \\
2MASSJ11040425-7639328  &  CHSM1715                &  192.31  &  -10.62     & 1.3482      &   \nodata \\
2MASSJ11040909-7627193  &  CT-Cha-A                &  191.78  &  -6.44      & 49.8157     &   \nodata \\
2MASSJ11044258-7741571  &  ISO-52                  &  193.15  &  -10.34     & 2.0116      &   na \\
2MASSJ11045701-7715569  &  T16                     &  194.46  &  -7.54      & 1.2572      &   \nodata \\
2MASSJ11062554-7633418  &  ESOHalpha559            &  209.30  &  -10.48     & 26.5032     &   \nodata \\
2MASSJ11063276-7625210  &  CHSM7869                &  187.14  &  -10.87     & $<$ 0.0723  &   \nodata \\
2MASSJ11064180-7635489  &  Hn-5                    &  195.29  &  -9.04      & 0.4821      &   \nodata \\
2MASSJ11064510-7727023  &  CHXR20                  &  185.47  &  -8.50      & $<$ 0.3644  &   \nodata \\
2MASSJ11065906-7718535  &  T23                     &  190.34  &  -7.95      & 11.5032     &   \nodata \\
2MASSJ11065939-7530559  &  2MJ11065939-7530559     &  196.24  &  -10.87     & 1.5752      &   \nodata \\
2MASSJ11071181-7625501  &  CHSM9484                &  199.48  &  -11.58     & $<$ 0.0822  &   na \\
2MASSJ11071206-7632232  &  T24                     &  195.75  &  -8.20      & 2.1143      &   \nodata \\
2MASSJ11071330-7743498  &  CHXR22E                 &  173.47  &  -10.81     & $<$ 0.3188  &   na,td \\
2MASSJ11071860-7732516  &  Cha-Ha-9                &  198.58  &  -10.67     & 0.4760      &   \nodata \\
2MASSJ11072074-7738073  &  Sz19                    &  190.62  &  -7.45      & 12.3622     &   \nodata \\
2MASSJ11072825-7652118  &  T27                     &  190.00  &  -8.13      & $<$ 0.3824  &   \nodata \\
2MASSJ11074245-7733593  &  ChaHalpha2              &  190.00  &  -9.84      & 1.1201      &   \nodata \\
2MASSJ11074366-7739411  &  T28                     &  194.81  &  -7.65      & 52.5996     &   \nodata \\
2MASSJ11074656-7615174  &  CHSM10862               &  194.22  &  -11.76     & 1.0675      &   na \\
2MASSJ11075792-7738449  &  Sz-22                   &  163.19  &  -8.31      & 6.8730      &   \nodata \\
2MASSJ11075809-7742413  &  T30                     &  184.46  &  -8.11      & 2.8416      &   \nodata \\
2MASSJ11080002-7717304  &  CHXR30A                 &  190.00  &  -9.93      & $<$ 0.3737  &   na \\
2MASSJ11080148-7742288  &  VW-Cha                  &  190.00  &  -7.42      & 20.8577     &   \nodata \\
2MASSJ11080297-7738425  &  ESO-Ha-562              &  190.00  &  -8.98      & 47.7823     &   \nodata \\
2MASSJ11081509-7733531  &  T33A                    &  190.00  &  -8.79      & 97.5590     &   \nodata \\
2MASSJ11081850-7730408  &  ISO138                  &  185.65  &  -11.63     & $<$ 0.0712  &   na \\
2MASSJ11082238-7730277  &  ISO-143                 &  193.28  &  -9.86      & $<$ 0.0772  &   \nodata \\
2MASSJ11082650-7715550  &  ISO147                  &  200.00  &  -10.89     & $<$ 0.0826  &   \nodata \\
2MASSJ11083905-7716042  &  Sz27                    &  188.36  &  -8.64      & 6.4824      &   td \\
2MASSJ11083952-7734166  &  Cha-Ha6                 &  179.32  &  -10.17     & $<$ 0.0664  &   \nodata \\
2MASSJ11085090-7625135  &  T37                     &  192.77  &  -10.53     & $<$ 0.0768  &   \nodata \\
2MASSJ11085367-7521359  &  2MJ11085367-7521359     &  188.27  &  -7.91      & 11.5165     &   \nodata \\
2MASSJ11085464-7702129  &  T38                     &  186.01  &  -9.07      & 1.7817      &   \nodata \\
2MASSJ11085497-7632410  &  ISO165                  &  194.65  &  -10.43     & $<$ 0.0783  &   \nodata \\
2MASSJ11091812-7630292  &  CHXR79                  &  187.63  &  -8.83      & $<$ 0.3645  &   \nodata \\
2MASSJ11092266-7634320  &  C1-6                    &  203.25  &  -9.21      & 2.0788      &   \nodata \\
2MASSJ11092379-7623207  &  T40                     &  192.30  &  -7.08      & 58.8462     &   \nodata \\
2MASSJ11094621-7634463  &  Hn10e                   &  195.04  &  -9.24      & 2.3551      &   \nodata \\
2MASSJ11094742-7726290  &  ISO207                  &  192.96  &  -8.93      & 71.2352     &   \nodata \\
2MASSJ11095215-7639128  &  ISO217                  &  240.14  &  -10.25     & $<$ 0.1191  &   \nodata \\
2MASSJ11095336-7728365  &  ISO220                  &  186.29  &  -10.27     & $<$ 0.0717  &   \nodata \\
2MASSJ11095340-7634255  &  Sz32                    &  201.96  &  -6.75      & 40.9535     &   \nodata \\
2MASSJ11095407-7629253  &  Sz33                    &  212.11  &  -8.94      & 17.9838     &   \nodata \\
2MASSJ11095873-7737088  &  T45                     &  191.29  &  -6.70      & 9.8888      &   \nodata \\
2MASSJ11100010-7634578  &  T44                     &  192.08  &  -6.50      & 658.7556    &   \nodata \\
2MASSJ11100369-7633291  &  Hn11                    &  201.01  &  -9.29      & 5.2263      &   \nodata \\
2MASSJ11100469-7635452  &  T45a                    &  195.00  &  -9.54      & 3.8179      &   \nodata \\
2MASSJ11100704-7629376  &  T46                     &  179.61  &  -8.53      & 3.0229      &   \nodata \\
2MASSJ11100785-7727480  &  ISO235                  &  199.93  &  -10.70     & $<$ 0.0826  &   \nodata \\
2MASSJ11101141-7635292  &  ISO-237                 &  195.37  &  -9.48      & 36.5996     &   na \\
Continues next page \\
\hline 
\end{tabular} 
\tablefoot{Stellar properties adapted from \citet{manara16a,manara17a} and disk masses from \citet{pascucci16} using the Gaia DR2 \citep{gaiadr2} distances. Notes: na = non-accretor, sl=sub-luminous, td = transition disk. } 
\end{center} 
\end{table*}

\begin{table*}  
\begin{center} 
\footnotesize 
\caption{\label{tab::chaI_bis} (End of) Stellar and disk masses for the Chamaeleon~I targets used here } 
\begin{tabular}{l|l| ccc|l   } 
\hline \hline 
2MASS Name   &   Other names    & dist    &    $\log\dot{M}_{\rm acc}$  &         $M_{\rm dust}$   &      Notes \\
 &            &             [pc]  &   [$M_\odot$/yr]            & [$M_\oplus$] & \\
\hline
2MASSJ11103801-7732399  &  CHXR47                  &  190.00  &  -7.89      & 2.5077      &   \nodata \\
2MASSJ11104141-7720480  &  ISO252                  &  204.22  &  -9.63      & $<$ 0.0861  &   \nodata \\
2MASSJ11104959-7717517  &  Sz37                    &  185.16  &  -7.60      & 26.1129     &   \nodata \\
2MASSJ11105333-7634319  &  T48                     &  194.69  &  -7.70      & 15.1512     &   \nodata \\
2MASSJ11105359-7725004  &  ISO256                  &  195.78  &  -10.09     & 3.9382      &   \nodata \\
2MASSJ11105597-7645325  &  Hn13                    &  190.00  &  -9.40      & 1.0454      &   \nodata \\
2MASSJ11113965-7620152  &  T49                     &  190.51  &  -7.18      & 10.2706     &   \nodata \\
2MASSJ11114632-7620092  &  CHX18N                  &  192.52  &  -7.90      & 17.0103     &   \nodata \\
2MASSJ11120351-7726009  &  ISO282                  &  185.49  &  -9.71      & 1.3134      &   \nodata \\
2MASSJ11120984-7634366  &  T50                     &  193.23  &  -9.11      & 2.1573      &   \nodata \\
2MASSJ11122441-7637064  &  T51                     &  193.81  &  -7.97      & $<$ 0.3800  &   \nodata \\
2MASSJ11122772-7644223  &  T52                     &  193.24  &  -7.31      & 28.4416     &   \nodata \\
2MASSJ11123092-7644241  &  CWCha                   &  196.00  &  -7.74      & 6.1133      &   \nodata \\
2MASSJ11124268-7722230  &  T54A                    &  201.58  &  -9.37      & $<$ 0.4207  &   na,td \\
2MASSJ11124861-7647066  &  Hn17                    &  191.24  &  -9.47      & $<$ 0.0755  &   \nodata \\
2MASSJ11132446-7629227  &  Hn18                    &  189.52  &  -9.58      & 3.7763      &   \nodata \\
2MASSJ11142454-7733062  &  Hn21W                   &  188.95  &  -8.82      & 3.5031      &   \nodata \\
2MASSJ11173700-7704381  &  Sz45                    &  188.38  &  -7.85      & 12.9369     &   td \\
2MASSJ11183572-7935548  &  2MJ11183572-7935548     &  94.62   &  -9.57      & 1.7129      &   td \\
2MASSJ11241186-7630425  &  2MJ11241186-7630425     &  184.75  &  -10.43     & 0.6530      &   td \\
2MASSJ11432669-7804454  &  2MJ11432669-7804454     &  190.00  &  -8.50      & 0.6446      &   \nodata \\
\hline 
\end{tabular} 
\tablefoot{Stellar properties adapted from \citet{manara16a,manara17a} and disk masses from \citet{pascucci16} using the Gaia DR2 \citep{gaiadr2} distances. Notes: na = non-accretor, sl=sub-luminous, td = transition disk. } 
\end{center} 
\end{table*}

\begin{table*}  
\begin{center} 
\footnotesize 
\caption{\label{tab::tds} Stellar and disk masses for the transition disks used here } 
\begin{tabular}{l|cc| ccc|l   } 
\hline \hline 
 Object  &    RA       &     DEC     &      dist [pc] &  $M_{\rm dust}$[$M_\oplus$] &       $\log\dot{M}_{\rm acc}$ &  Ref \\
\hline 
 J16083070   & 16:08:30.68  &  -38:28:27.22 &   156     &       34.45    &       -9.32     & \citealt{alcala17}  \\
 RYLup       & 15:59:28.37  &  -40:21:51.58 &    159       &    72.84    &       -8.16     & \citealt{alcala17}  \\
 Sz111       & 16:08:54.67  &  -39:37:43.49 &    158       &    46.97    &       -9.47     & \citealt{alcala14}  \\
 Sz100       & 16:08:25.75  &  -39:06:01.64 &    137       &    10.74    &       -9.87     & \citealt{alcala14}  \\
 Sz118       & 16:09:48.64  &  -39:11:17.24 &    164       &    17.77    &       -9.21     & \citealt{alcala17}  \\
 Sz123A      & 16:10:51.57  &  -38:53:14.10 &     158      &    10.66    &       -9.21     & \citealt{alcala14}  \\
J15583692     & 15:58:36.90  &  -22:57:15.57 &    166   &       47.26   &       -7.80      & Manara et al. in prep.  \\
J16042165     & 16:04:21.64  &  -21:30:28.98 &    150   &       42.66   &       -10.51     & Manara et al. in prep. \\
 SzCha          & 10:58:16.71  &  -77:17:17.15 &         190       &    147.41   &       -7.65     & \citealt{manara14}  \\
 J10563044   & 10:56:30.31  &  -77:11:39.25 &    183       &    50.94    &       -9.22     & \citealt{manara16a}  \\
DoAr44        & 16:31:33.46  &  -24:27:37.52 &   146    &       50.88   &       -8.12      & \citealt{manara14}  \\
HD100546      & 11:33:25.36  &  -70:11:41.27 &   110    &       152.72  &       -6.84      & \citealt{fairlamb15}  \\
HD135344B     & 15:15:48.42  &  -37:09:16.33 &   135    &       110.93  &       -7.33      & \citealt{fairlamb15}  \\
LkCa15        & 04:39:17.80  &  +22:21:03.22 &    159   &       144.40  &       -8.31      & \citealt{manara14}  \\
SR21          & 16:27:10.27  &  -24:19:13.01 &    138   &       82.29   &       -7.81      & \citealt{manara14}  \\
SR24S         & 16:26:58.50  &  -24:45:37.20 &    115   &       82.48   &       -7.47      &  \citealt{natta06}  \\
 Sz91        & 16:07:11.57  &  -39:03:47.85 &      159     &    11.31    &       -9.07     & \citealt{alcala14}  \\
TCha          & 11:57:13.28  &  -79:21:31.72 &    110   &       35.54   &       -8.40      &  \citealt{schisano09}  \\
HD34282       & 05:16:00.48  &  -09:48:35.42 &    311   &       344.39  &       -8.30      &  \citealt{fairlamb15}  \\
CIDA1         & 04:14:17.62  &  +28:06:09.28 &    136   &       7.46    &       -7.92      & \citealt{pinilla18_tds} \\
CQTau         & 05:35:58.47  &  +24:44:53.70 &    163   &       105.71  &       $<$-7.94   &    \citealt{mendigutia11}  \\
HD142666      & 15:56:40.2  &   -22:01:39.5  &    148   &       80.79   &       $<$-8.27   &     \citealt{fairlamb15}       \\
RYTau         & 04:21:57.42  &  +28:26:35.13 &    130   &       75.03   &       -7.65      & \citealt{mendigutia11}   \\
UXTauA        & 04:30:04.00  &  +18:13:49.18 &    140   &       30.16   &       -8.71      & \citealt{rigliaco15}  \\
RXJ1615       & 16:15:20.23  &  -32:55:05.36 &   158    &       209.81     &     -8.49      & \citealt{manara14}  \\
DMTau         & 04:33:48.75  &  +18:10:09.66 &   145    &       25.22     &      -8.02      & \citealt{manara14}  \\
PDS70   & 14:08:10.15   & -41:23:52.58 & 113    &       39.79   & -10.26        & \citealt{haffert19} \\
\hline 
\end{tabular} 
\tablefoot{References are for the stellar and accretion properties. All disk masses from Lupus are from \citet{ansdell16}, that for PDS70 from \citet{keppler19}, for DMTau from \citet{kudo18}, the others from \citet{pinilla18_tds}. } 
\end{center} 
\end{table*}

\end{document}